\documentclass[RNAAS]{aastex62}

\received{Sep 23, 2018}
\revised{Sep 24, 2018}
\accepted{Sep 24, 2018}
\submitjournal{ApJ}

\shorttitle{Update on the GJ1046 system}
\shortauthors{Trifonov et al.}

\begin{document}

\title{New HARPS and FEROS observations of GJ1046.}

\correspondingauthor{Trifon Trifonov}
\email{trifonov@mpia.de}

\author{Trifon Trifonov}
\affil{Max-Planck-Institut f\"{u}r Astronomie, K\"{o}nigstuhl  17, 69117 Heidelberg, Germany}

\author{Martin K\"{u}rster}
\affiliation{Max-Planck-Institut f\"{u}r Astronomie, K\"{o}nigstuhl  17, 69117 Heidelberg, Germany}

\author{Sabine Reffert}
\affiliation{Landessternwarte, Zentrum f\"{u}r Astronomie der Universit\"{a}t Heidelberg, K\"{o}nigstuhl 12, 69117 Heidelberg, Germany}

\author{Mathias Zechmeister}
\affiliation{Institut f\"{u}r Astrophysik, Georg-August-Universit\"{a}t, Friedrich-Hund-Platz 1, 37077, G\"{o}ttingen, Germany}

\author{Michael Endl}
\affiliation{McDonald Observatory and Department of Astronomy, The University of Texas at Austin, Austin, TX 78712, USA}

\author{Florian Rodler} 
\affil{European Southern Observatory (ESO), Alonso de Cordova 3107, Vitacura, Santiago de Chile, Chile}

\author{Davide Gandolfi}
\affil{Dipartimento di Fisica, Universit\`{a} di Torino, via P. Giuria 1, 10125 Torino, Italy}

\author{Oscar Barrag\'{a}n}
\affil{Dipartimento di Fisica, Universit\`{a} di Torino, via P. Giuria 1, 10125 Torino, Italy}

\author{Thomas Henning}
\affil{Max-Planck-Institut f\"{u}r Astronomie, K\"{o}nigstuhl  17, 69117 Heidelberg, Germany}

\author{Man Hoi Lee}
\affil{Department of Earth Sciences, The University of Hong Kong, Pokfulam Road, Hong Kong}
\affil{Department of Physics, The University of Hong Kong, Pokfulman Road, Hong Kong}

\author{Olga Zakhozhay}
\affil{Max-Planck-Institut f\"{u}r Astronomie, K\"{o}nigstuhl  17, 69117 Heidelberg, Germany}
\affil{Main Astronomical Observatory, National Academy of Sciences of the Ukraine, Ukraine}

\author{Paula Sarkis}
\affil{Max-Planck-Institut f\"{u}r Astronomie, K\"{o}nigstuhl  17, 69117 Heidelberg, Germany}

\author{Paul Heeren}
\affiliation{Landessternwarte, Zentrum f\"{u}r Astronomie der Universit\"{a}t Heidelberg, K\"{o}nigstuhl 12, 69117 Heidelberg, Germany}

\author{Marcelo Tala}
\affiliation{Landessternwarte, Zentrum f\"{u}r Astronomie der Universit\"{a}t Heidelberg, K\"{o}nigstuhl 12, 69117 Heidelberg, Germany}

\author{Vera Wolthoff}
\affiliation{Landessternwarte, Zentrum f\"{u}r Astronomie der Universit\"{a}t Heidelberg, K\"{o}nigstuhl 12, 69117 Heidelberg, Germany}
 
\author{Stefan S. Brems}
\affiliation{Landessternwarte, Zentrum f\"{u}r Astronomie der Universit\"{a}t Heidelberg, K\"{o}nigstuhl 12, 69117 Heidelberg, Germany}

\author{Stephan Stock}
\affiliation{Landessternwarte, Zentrum f\"{u}r Astronomie der Universit\"{a}t Heidelberg, K\"{o}nigstuhl 12, 69117 Heidelberg, Germany}

\author{Angela Hempel}
\affiliation{Departemento de Ciencias Fisicas, Universidad Andres Bello, Campus Casona de Las Condes, Fern\`{a}ndez Concha 700, Santiago, Chile }
\affil{Max-Planck-Institut f\"{u}r Astronomie, K\"{o}nigstuhl  17, 69117 Heidelberg, Germany}

\author{Diana Kossakowski}
\affil{Max-Planck-Institut f\"{u}r Astronomie, K\"{o}nigstuhl  17, 69117 Heidelberg, Germany}

\keywords{Techniques: radial velocities $-$ Planets and satellites: detection }

\section{Introduction} \label{sec:intro}

GJ1046 is a M2.5\,V, $V$=11.62 mag star reported to host 
a substellar companion with a minimum mass in the brown dwarf mass regime \citep{Kuerster2008}. 
The system is peculiar because of the short orbital period of only $\sim$169 days,
which locates the companion in the so-called ``brown dwarf`` desert \citep{Marcy2000} around a low-mass star. 
GJ1046 b was discovered with only 14 high precision Doppler measurements 
taken with UVES \citep[VLT-UT2;][]{Dekker2000}, which, given the large RV semi-amplitude of $\sim$1830 ms$^{-1}$, 
were sufficient to constrain the orbit.
Yet, the somewhat sparse phase coverage of the orbit may lead to model ambiguity, 
such as two near resonant companions masking as one eccentric orbit \citep[see][]{Escude2010,Wittenmyer2013,Kurster2015},
or an orbital period that could in fact be approximately twice the reported one \citep[see Fig.1 in][]{Kuerster2008}.
Therefore, we have continued observing GJ1046 over the years in an attempt to constrain the orbit better.

In this paper we present new precise Doppler data of GJ1046 taken between November 2005 and July 2018
with the HARPS \citep[][]{Mayor2003} and FEROS \citep[][]{Kaufer1999}
high-resolution spectographs. 
In addition, we provide a new stellar mass estimate of GJ1046 and we update the orbital 
parameters of the GJ1046 system. 
These new data and analysis could be used together with the GAIA epoch astrometry, when available,
for breaking the $\sin i$ degeneracy and revealing the true mass of the GJ1046 system.

\section{HARPS and FEROS observations and data reduction} \label{sec:observations}

GJ1046 is part of our HARPS and FEROS RV monitoring program of a sample of 36 southern stars, 
which are known to host a single moderately eccentric sub-stellar companion \citep{Trifonov2017}. 
Given the published UVES data of GJ1046 we were unable to completely refute
the existence of a second companion, and thus we decided to obtain more Doppler measurements for this target.
We did not find a second companion for GJ1046, but our RV data are valuable to 
determine accurately the spectroscopic orbital properties of the system.

We process the FEROS spectra using the \textit{CERES} pipeline \citep{Brahm2017a}, 
whereas the HARPS spectra were re-processed using \textit{SERVAL} \citep[][]{Zechmeister2017},
which is shown to provide a better precision than the ESO-DRS pipeline for M dwarfs  
\citep[e.g. see][]{Trifonov2018a,Kaminski2018}.
We reduced a total of 34 FEROS and 27 HARPS spectra, achieving a mean RV 
precision of 14.5 ms$^{-1}$ and 1.8 ms$^{-1}$, respectively.
However, we split the HARPS data into two separate temporal subsets,  
HARPS and HARPS+, pertinent to before and after the HARPS fibre upgrade
in June 2015, respectively, which introduced an RV offset  
\citep[][]{LoCurto2015}.

\section{New orbital and Stellar mass estimates} \label{sec:fitting}

We model the available Doppler data of GJ1046 using a downhill Simplex algorithm \citep[][]{NelderMead,Press},
which optimizes simultaneously the parameters of a Keplerian model, the RV data offsets, and the 
RV data jitter noise \citep{Baluev2009}. 
We estimate the orbital parameter uncertainties of the best fit 
by adopting the 1$\sigma$ confidence levels of the parameter posterior distribution 
sampled with the $emcee$ MCMC sampler \citep{emcee}.
Our best fit yields the following values:
Semi-amplitude K$_{\rm b}$ = 1827.0$_{-1.2}^{+2.4}$ ms$^{-1}$,
period P$_{\rm b}$ = 168.844$_{-0.005}^{+0.001}$ days, eccentricity 
$e_{\rm b}$ = 0.2788$_{-0.001}^{+0.001}$, argument of periastron
$\omega_{\rm b}$ = 92.24$_{-0.14}^{+0.39}$ deg, and time of periastron passage 
$t_{\omega_{\rm b}}$ = 2453225.516$_{-0.120}^{+0.119}$ BJD.
These estimates are consistent with those in \citet{Kuerster2008}, but the uncertainties are considerably smaller.


To estimate the stellar mass of GJ1046 we constructed a new mass-luminosity relation by combining the data originally used by \citet{Benedict2016} and \citet{Delfosse2000} for their K-band mass-luminostiy relations.
We also included the Sun and TRAPPIST-1 \citep{Gillon2016} K-band-masses in order to constrain the high and the low ends of the relation.
The best-fit polynomial of the type $M=a_{\rm 0}+a_{\rm 1}K+a_{\rm 4}K^4$ has the following coefficients; $a_0$ = 1.735$\pm$0.005, 
$a_1$ = $-$0.2262$\pm$0.0016, $a_4$ = 0.000062$\pm$0.0000016.
With an estimated distance from the Gaia DR2 parallax of $d$ = 15.19$_{-0.03}^{+0.03}$ pc \citep{Bailer_Jones2018} 
and an absolute K-magnitude of K$_{\rm abs}$ = 6.12$\pm$0.02 mag, the resulting mass of GJ1046 is $M$ = 0.437$\pm$0.012 $M_\odot$, yielding an orbital semi-major axis of $a_{\rm b}$ = 0.463$\pm$0.004 au and a 
minimum companion mass $m \sin i_{\rm b}$ = 28.62$\pm$0.51 M$_{\rm Jup}$.

\section{Conclusions} \label{sec:Conclusions}

GJ1046 b has a minimum mass consistent with a brown dwarf companion and a
very small orbital separation of 0.463 au. 
Only a small number of similar objects are known around M dwarfs.
However, the possibility that the substellar companion could be of a stellar nature still remains. 
As in \citet{Kuerster2008} we tried to fit the astrometric orbit to the Hipparcos data taking the more accurate 
parallax and proper motions from TGAS, but could not significantly improve on
the known range of possible inclinations.
With the release of the GAIA epoch astrometry and the Doppler data presented here this degeneracy should be resolved.

\begin{figure*}[]
\begin{center}$
\begin{array}{ccc}

\includegraphics[width=18cm]{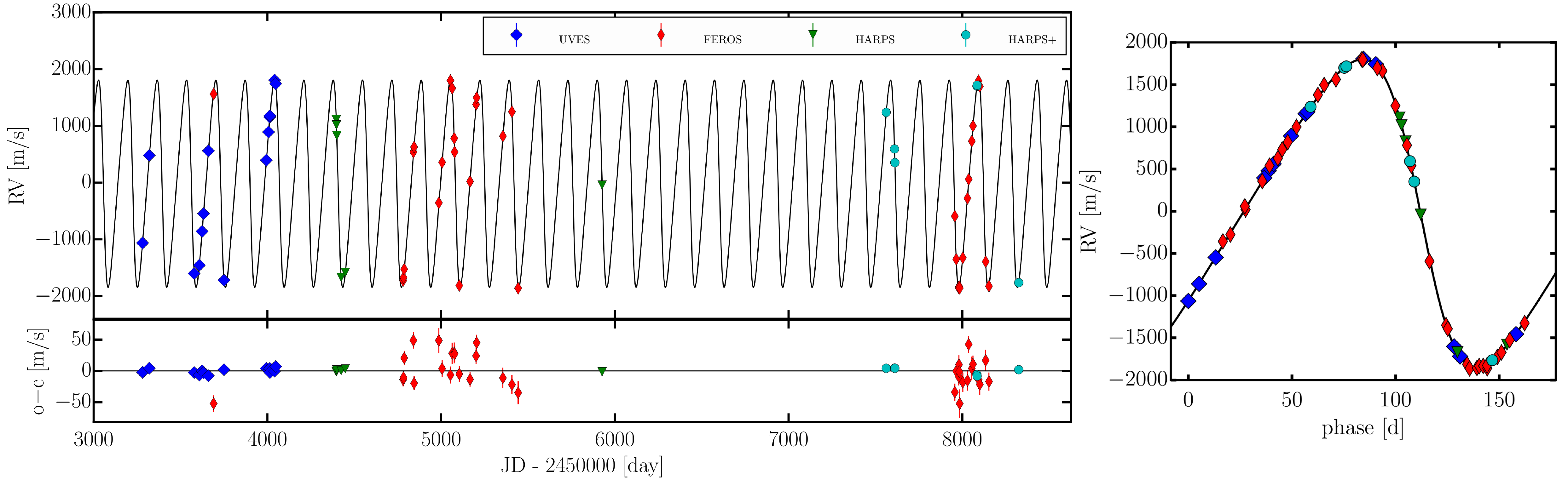}
 
\end{array} $
\end{center}

\caption{Archival and new precise Doppler measurements of GJ1046. The left panel shows time series data from UVES, 
FEROS, HARPS and HARPS+  fitted with a Keplerian model. The right panel shows the phase-folded model and RVs at the best fit period.
FEROS RVs have larger scatter when compared with UVES and HARPS, but are adequate for the determination of the orbit. 
}   
\label{res_angle_lib_dist} 
\end{figure*}

\acknowledgments
 
This research is based on observations collected at 
the European Organization for Astronomical Research in the Southern Hemisphere under ESO 
programmes 60.A-9700, 180.C-0886, 183.C-0437, 097.C-0090, 0100.C-0414 and MPG programmes 076.A-9005, 082.A-9006, 083.A-9012, 
084.A-9004, 085.A-9009, 099.A-9009, 0100.A-9006.
S.S.\ and S.R.\ acknowledge support by the DFG Research Unit FOR~2544 {\it Blue  Planets around Red Stars}, project no.~RE~2694/4-1. V.W.\ and S.R.\ further acknowledge support by the DFG Priority Program SPP~1992 {\it Exploring the Diversity of Extrasolar Planets} (RE~2694/5-1). M.H.L. is supported in part by Hong Kong
 RGC grant HKU 17305015.

\bibliographystyle{aasjournal}

\bibliography{gj1046_bib}

\begin{thebibliography}{}
\expandafter\ifx\csname natexlab\endcsname\relax\def\natexlab#1{#1}\fi
\providecommand{\url}[1]{\href{#1}{#1}}
\providecommand{\dodoi}[1]{doi:~\href{http://doi.org/#1}{\nolinkurl{#1}}}
\providecommand{\doeprint}[1]{\href{http://ascl.net/#1}{\nolinkurl{http://ascl.net/#1}}}
\providecommand{\doarXiv}[1]{\href{https://arxiv.org/abs/#1}{\nolinkurl{https://arxiv.org/abs/#1}}}

\bibitem[{{Anglada-Escud{\'e}} {et~al.}(2010){Anglada-Escud{\'e}},
  {L{\'o}pez-Morales}, \& {Chambers}}]{Escude2010}
{Anglada-Escud{\'e}}, G., {L{\'o}pez-Morales}, M., \& {Chambers}, J.~E. 2010,
  \apj, 709, 168, \dodoi{10.1088/0004-637X/709/1/168}

\bibitem[{{Bailer-Jones} {et~al.}(2018){Bailer-Jones}, {Rybizki}, {Fouesneau},
  {Mantelet}, \& {Andrae}}]{Bailer_Jones2018}
{Bailer-Jones}, C.~A.~L., {Rybizki}, J., {Fouesneau}, M., {Mantelet}, G., \&
  {Andrae}, R. 2018, ArXiv e-prints.
\newblock \doarXiv{1804.10121}

\bibitem[{{Baluev}(2009)}]{Baluev2009}
{Baluev}, R.~V. 2009, \mnras, 393, 969,
  \dodoi{10.1111/j.1365-2966.2008.14217.x}

\bibitem[{{Benedict} {et~al.}(2016){Benedict}, {Henry}, {Franz}, {McArthur},
  {Wasserman}, {Jao}, {Cargile}, {Dieterich}, {Bradley}, {Nelan}, \&
  {Whipple}}]{Benedict2016}
{Benedict}, G.~F., {Henry}, T.~J., {Franz}, O.~G., {et~al.} 2016, \aj, 152,
  141, \dodoi{10.3847/0004-6256/152/5/141}

\bibitem[{{Brahm} {et~al.}(2017){Brahm}, {Jord{\'a}n}, \&
  {Espinoza}}]{Brahm2017a}
{Brahm}, R., {Jord{\'a}n}, A., \& {Espinoza}, N. 2017, \pasp, 129, 034002,
  \dodoi{10.1088/1538-3873/aa5455}

\bibitem[{{Dekker} {et~al.}(2000){Dekker}, {D'Odorico}, {Kaufer}, {Delabre}, \&
  {Kotzlowski}}]{Dekker2000}
{Dekker}, H., {D'Odorico}, S., {Kaufer}, A., {Delabre}, B., \& {Kotzlowski}, H.
  2000, in \procspie, Vol. 4008, Optical and IR Telescope Instrumentation and
  Detectors, ed. M.~{Iye} \& A.~F. {Moorwood}, 534--545

\bibitem[{{Delfosse} {et~al.}(2000){Delfosse}, {Forveille}, {S{\'e}gransan},
  {Beuzit}, {Udry}, {Perrier}, \& {Mayor}}]{Delfosse2000}
{Delfosse}, X., {Forveille}, T., {S{\'e}gransan}, D., {et~al.} 2000, \aap, 364,
  217

\bibitem[{{Foreman-Mackey} {et~al.}(2013){Foreman-Mackey}, {Hogg}, {Lang}, \&
  {Goodman}}]{emcee}
{Foreman-Mackey}, D., {Hogg}, D.~W., {Lang}, D., \& {Goodman}, J. 2013, PASP,
  125, 306, \dodoi{10.1086/670067}

\bibitem[{{Gillon} {et~al.}(2016){Gillon}, {Jehin}, {Lederer}, {Delrez}, {de
  Wit}, {Burdanov}, {Van Grootel}, {Burgasser}, {Triaud}, {Opitom}, {Demory},
  {Sahu}, {Bardalez Gagliuffi}, {Magain}, \& {Queloz}}]{Gillon2016}
{Gillon}, M., {Jehin}, E., {Lederer}, S.~M., {et~al.} 2016, \nat, 533, 221,
  \dodoi{10.1038/nature17448}

\bibitem[{{Kaminski} {et~al.}(2018){Kaminski}, {Trifonov}, {Caballero},
  {Quirrenbach}, {Ribas}, {Reiners}, {Amado}, {Zechmeister}, {Dreizler},
  {Perger}, {Tal-Or}, {Bonfils}, {Mayor}, {Astudillo-Defru}, {Bauer},
  {B{\'e}jar}, {Cifuentes}, {Colom{\'e}}, {Cort{\'e}s-Contreras}, {Delfosse},
  {D{\'{\i}}ez-Alonso}, {Forveille}, {Guenther}, {Hatzes}, {Henning},
  {Jeffers}, {K{\"u}rster}, {Lafarga}, {Luque}, {Mandel}, {Montes}, {Morales},
  {Passegger}, {Pedraz}, {Reffert}, {Sadegi}, {Schweitzer}, {Seifert}, {Stahl},
  \& {Udry}}]{Kaminski2018}
{Kaminski}, A., {Trifonov}, T., {Caballero}, J.~A., {et~al.} 2018, ArXiv
  e-prints.
\newblock \doarXiv{1808.01183}

\bibitem[{{Kaufer} {et~al.}(1999){Kaufer}, {Stahl}, {Tubbesing},
  {N{\o}rregaard}, {Avila}, {Francois}, {Pasquini}, \& {Pizzella}}]{Kaufer1999}
{Kaufer}, A., {Stahl}, O., {Tubbesing}, S., {et~al.} 1999, The Messenger, 95, 8

\bibitem[{{K{\"u}rster} {et~al.}(2008){K{\"u}rster}, {Endl}, \&
  {Reffert}}]{Kuerster2008}
{K{\"u}rster}, M., {Endl}, M., \& {Reffert}, S. 2008, \aap, 483, 869,
  \dodoi{10.1051/0004-6361:200809419}

\bibitem[{{K{\"u}rster} {et~al.}(2015){K{\"u}rster}, {Trifonov}, {Reffert},
  {Kostogryz}, \& {Rodler}}]{Kurster2015}
{K{\"u}rster}, M., {Trifonov}, T., {Reffert}, S., {Kostogryz}, N.~M., \&
  {Rodler}, F. 2015, \aap, 577, A103, \dodoi{10.1051/0004-6361/201525872}

\bibitem[{{Lo Curto} {et~al.}(2015){Lo Curto}, {Pepe}, {Avila}, {Boffin},
  {Bovay}, {Chazelas}, {Coffinet}, {Fleury}, {Hughes}, {Lovis}, {Maire},
  {Manescau}, {Pasquini}, {Rihs}, {Sinclaire}, \& {Udry}}]{LoCurto2015}
{Lo Curto}, G., {Pepe}, F., {Avila}, G., {et~al.} 2015, The Messenger, 162, 9

\bibitem[{{Marcy} \& {Butler}(2000)}]{Marcy2000}
{Marcy}, G.~W., \& {Butler}, R.~P. 2000, \pasp, 112, 137,
  \dodoi{10.1086/316516}

\bibitem[{{Mayor} {et~al.}(2003){Mayor}, {Pepe}, {Queloz}, {Bouchy},
  {Rupprecht}, {Lo Curto}, {Avila}, {Benz}, {Bertaux}, {Bonfils}, {Dall},
  {Dekker}, {Delabre}, {Eckert}, {Fleury}, {Gilliotte}, {Gojak}, {Guzman},
  {Kohler}, {Lizon}, {Longinotti}, {Lovis}, {Megevand}, {Pasquini}, {Reyes},
  {Sivan}, {Sosnowska}, {Soto}, {Udry}, {van Kesteren}, {Weber}, \&
  {Weilenmann}}]{Mayor2003}
{Mayor}, M., {Pepe}, F., {Queloz}, D., {et~al.} 2003, The Messenger, 114, 20

\bibitem[{Nelder \& Mead(1965)}]{NelderMead}
Nelder, J.~A., \& Mead, R. 1965, Computer Journal, 7, 308

\bibitem[{{Press} {et~al.}(1992){Press}, {Teukolsky}, {Vetterling}, \&
  {Flannery}}]{Press}
{Press}, W.~H., {Teukolsky}, S.~A., {Vetterling}, W.~T., \& {Flannery}, B.~P.
  1992, {Numerical recipes in FORTRAN. The art of scientific computing}

\bibitem[{{Trifonov} {et~al.}(2017){Trifonov}, {K{\"u}rster}, {Zechmeister},
  {Zakhozhay}, {Reffert}, {Lee}, {Rodler}, {Vogt}, \& {Brems}}]{Trifonov2017}
{Trifonov}, T., {K{\"u}rster}, M., {Zechmeister}, M., {et~al.} 2017, \aap, 602,
  L8, \dodoi{10.1051/0004-6361/201731044}

\bibitem[{{Trifonov} {et~al.}(2018){Trifonov}, {K{\"u}rster}, {Zechmeister},
  {Tal-Or}, {Caballero}, {Quirrenbach}, {Amado}, {Ribas}, {Reiners}, {Reffert},
  {Dreizler}, {Hatzes}, {Kaminski}, {Launhardt}, {Henning}, {Montes},
  {B{\'e}jar}, {Mundt}, {Pavlov}, {Schmitt}, {Seifert}, {Morales}, {Nowak},
  {Jeffers}, {Rodr{\'{\i}}guez-L{\'o}pez}, {del Burgo}, {Anglada-Escud{\'e}},
  {L{\'o}pez-Santiago}, {Mathar}, {Ammler-von Eiff}, {Guenther}, {Barrado},
  {Gonz{\'a}lez Hern{\'a}ndez}, {Mancini}, {St{\"u}rmer}, {Abril}, {Aceituno},
  {Alonso-Floriano}, {Antona}, {Anwand-Heerwart}, {Arroyo-Torres}, {Azzaro},
  {Baroch}, {Bauer}, {Becerril}, {Ben{\'{\i}}tez}, {Berdi{\~n}as}, {Bergond},
  {Bl{\"u}mcke}, {Brinkm{\"o}ller}, {Cano}, {C{\'a}rdenas V{\'a}zquez},
  {Casal}, {Cifuentes}, {Claret}, {Colom{\'e}}, {Cort{\'e}s-Contreras},
  {Czesla}, {D{\'{\i}}ez-Alonso}, {Feiz}, {Fern{\'a}ndez}, {Ferro},
  {Fuhrmeister}, {Galad{\'{\i}}-Enr{\'{\i}}quez}, {Garcia-Piquer},
  {Garc{\'{\i}}a Vargas}, {Gesa}, {G{\'o}mez Galera}, {Gonz{\'a}lez-Peinado},
  {Gr{\"o}zinger}, {Grohnert}, {Gu{\`a}rdia}, {Guijarro}, {de Guindos},
  {Guti{\'e}rrez-Soto}, {Hagen}, {Hauschildt}, {Hedrosa}, {Helmling},
  {Hermelo}, {Hern{\'a}ndez Arab{\'{\i}}}, {Hern{\'a}ndez Casta{\~n}o},
  {Hern{\'a}ndez Hernando}, {Herrero}, {Huber}, {Huke}, {Johnson}, {de Juan},
  {Kim}, {Klein}, {Kl{\"u}ter}, {Klutsch}, {Lafarga}, {Lamp{\'o}n}, {Lara},
  {Laun}, {Lemke}, {Lenzen}, {L{\'o}pez del Fresno}, {L{\'o}pez-Gonz{\'a}lez},
  {L{\'o}pez-Puertas}, {L{\'o}pez Salas}, {Luque}, {Mag{\'a}n Madinabeitia},
  {Mall}, {Mandel}, {Marfil}, {Mar{\'{\i}}n Molina}, {Maroto Fern{\'a}ndez},
  {Mart{\'{\i}}n}, {Mart{\'{\i}}n-Ruiz}, {Marvin}, {Mirabet}, {Moya},
  {Moreno-Raya}, {Nagel}, {Naranjo}, {Nortmann}, {Ofir}, {Oreiro}, {Pall{\'e}},
  {Panduro}, {Pascual}, {Passegger}, {Pedraz}, {P{\'e}rez-Calpena}, {P{\'e}rez
  Medialdea}, {Perger}, {Perryman}, {Pluto}, {Rabaza}, {Ram{\'o}n}, {Rebolo},
  {Redondo}, {Reinhardt}, {Rhode}, {Rix}, {Rodler}, {Rodr{\'{\i}}guez},
  {Rodr{\'{\i}}guez Trinidad}, {Rohloff}, {Rosich}, {Sadegi},
  {S{\'a}nchez-Blanco}, {S{\'a}nchez Carrasco}, {S{\'a}nchez-L{\'o}pez},
  {Sanz-Forcada}, {Sarkis}, {Sarmiento}, {Sch{\"a}fer}, {Schiller},
  {Sch{\"o}fer}, {Schweitzer}, {Solano}, {Stahl}, {Strachan}, {Su{\'a}rez},
  {Tabernero}, {Tala}, {Tulloch}, {Veredas}, {Vico Linares}, {Vilardell},
  {Wagner}, {Winkler}, {Wolthoff}, {Xu}, {Yan}, \& {Zapatero
  Osorio}}]{Trifonov2018a}
---. 2018, \aap, 609, A117, \dodoi{10.1051/0004-6361/201731442}

\bibitem[{{Wittenmyer} {et~al.}(2013){Wittenmyer}, {Wang}, {Horner}, {Tinney},
  {Butler}, {Jones}, {O'Toole}, {Bailey}, {Carter}, {Salter}, {Wright}, \&
  {Zhou}}]{Wittenmyer2013}
{Wittenmyer}, R.~A., {Wang}, S., {Horner}, J., {et~al.} 2013, \apjs, 208, 2,
  \dodoi{10.1088/0067-0049/208/1/2}

\bibitem[{{Zechmeister} {et~al.}(2018){Zechmeister}, {Reiners}, {Amado},
  {Azzaro}, {Bauer}, {B{\'e}jar}, {Caballero}, {Guenther}, {Hagen}, {Jeffers},
  {Kaminski}, {K{\"u}rster}, {Launhardt}, {Montes}, {Morales}, {Quirrenbach},
  {Reffert}, {Ribas}, {Seifert}, {Tal-Or}, \& {Wolthoff}}]{Zechmeister2017}
{Zechmeister}, M., {Reiners}, A., {Amado}, P.~J., {et~al.} 2018, \aap, 609,
  A12, \dodoi{10.1051/0004-6361/201731483}

\end{thebibliography}


\begin{table}
\caption{Doppler measurements of GJ 1046} 
\label{table:GJ 1046} 

\centering  

\begin{tabular}{c c c c } 

\hline\hline    
\noalign{\vskip 0.5mm}

Epoch [JD] & RV [m\,s$^{-1}$] & $\sigma_{RV}$  & instrument \\  

\hline     
\noalign{\vskip 0.5mm}    

2453690.718   &   1561.42   &    13.00   &  FEROS \\ 
2454396.742   &   1108.61   &    1.59   &  HARPS \\ 
2454397.609   &   1031.69   &    1.87   &  HARPS \\ 
2454397.870   &   1006.97   &    1.55   &  HARPS \\ 
2454399.673   &   827.51   &    1.65   &  HARPS \\ 
2454424.707   &   -1670.90   &    1.78   &  HARPS \\ 
2454448.629   &   -1582.85   &    2.90   &  HARPS \\ 
2454781.669   &   -1726.28   &    10.70   &  FEROS \\ 
2454783.659   &   -1672.98   &    10.10   &  FEROS \\ 
2454787.625   &   -1527.48   &    11.40   &  FEROS \\ 
2454840.601   &   539.12   &    13.00   &  FEROS \\ 
2454844.623   &   627.52   &    11.20   &  FEROS \\ 
2454986.924   &   -357.58   &    20.10   &  FEROS \\ 
2455005.944   &   356.32   &    13.10   &  FEROS \\ 
2455053.893   &   1799.62   &    13.50   &  FEROS \\ 
2455063.896   &   1662.32   &    16.80   &  FEROS \\ 
2455075.724   &   781.42   &    16.60   &  FEROS \\ 
2455077.857   &   538.92   &    13.00   &  FEROS \\ 
2455104.772   &   -1814.68   &    12.50   &  FEROS \\ 
2455166.730   &   18.92   &    11.60   &  FEROS \\ 
2455201.608   &   1377.72   &    12.70   &  FEROS \\ 
2455204.626   &   1495.52   &    12.80   &  FEROS \\ 
2455355.908   &   818.82   &    16.50   &  FEROS \\ 
2455407.855   &   1248.82   &    15.30   &  FEROS \\ 
2455443.630   &   -1862.18   &    18.30   &  FEROS \\ 
2455926.585   &   -40.96   &    1.73   &  HARPS \\ 
2457561.879   &   1236.85   &    1.77   &  HARPS+ \\ 
2457561.900   &   1235.40   &    1.36   &  HARPS+ \\ 
2457561.920   &   1235.66   &    1.28   &  HARPS+ \\ 
2457609.839   &   592.35   &    1.10   &  HARPS+ \\ 
2457609.860   &   590.40   &    1.07   &  HARPS+ \\ 
2457609.881   &   590.89   &    1.16   &  HARPS+ \\ 
2457611.850   &   351.43   &    1.15   &  HARPS+ \\ 
2457611.872   &   352.78   &    1.13   &  HARPS+ \\

\hline           
\end{tabular}

\end{table}

\begin{table}
\caption{Doppler measurements of GJ 1046 (continue)} 
\label{table:GJ 1046} 

\centering  

\begin{tabular}{c c c c } 

\hline\hline    
\noalign{\vskip 0.5mm}

Epoch [JD] & RV [m\,s$^{-1}$] & $\sigma_{RV}$  & instrument \\  

\hline     
\noalign{\vskip 0.5mm}

2457611.892   &   351.20   &    1.03   &  HARPS+ \\ 
2457611.933   &   344.97   &    1.20   &  HARPS+ \\ 
2457611.957   &   329.34   &    11.68   &  HARPS+ \\ 
2457956.924   &   -592.08   &    13.80   &  FEROS \\ 
2457964.933   &   -1350.88   &    13.30   &  FEROS \\ 
2457979.823   &   -1856.78   &    14.10   &  FEROS \\ 
2457980.911   &   -1834.98   &    14.70   &  FEROS \\ 
2457982.852   &   -1832.58   &    12.40   &  FEROS \\ 
2457984.809   &   -1859.88   &    22.70   &  FEROS \\ 
2458002.826   &   -1325.88   &    15.60   &  FEROS \\ 
2458029.742   &   -276.68   &    16.30   &  FEROS \\ 
2458036.696   &   59.02   &    13.40   &  FEROS \\ 
2458054.743   &   731.52   &    18.90   &  FEROS \\ 
2458061.627   &   998.92   &    13.60   &  FEROS \\ 
2458084.626   &   1697.03   &    1.38   &  HARPS+ \\ 
2458084.648   &   1699.84   &    1.34   &  HARPS+ \\ 
2458084.669   &   1694.68   &    1.56   &  HARPS+ \\ 
2458085.648   &   1713.41   &    1.58   &  HARPS+ \\ 
2458085.670   &   1714.37   &    1.27   &  HARPS+ \\ 
2458085.691   &   1716.77   &    1.27   &  HARPS+ \\ 
2458093.601   &   1792.52   &    15.10   &  FEROS \\ 
2458100.607   &   1696.22   &    16.90   &  FEROS \\ 
2458134.585   &   -1392.78   &    16.90   &  FEROS \\ 
2458153.546   &   -1826.28   &    14.60   &  FEROS \\ 
2458324.896   &   -1763.12   &    2.30   &  HARPS+ \\ 
2458324.917   &   -1766.98   &    1.55   &  HARPS+ \\ 
2458324.938   &   -1769.52   &    2.09   &  HARPS+ \\ 
  
\hline           
\end{tabular}

\tablecomments{The UVES RVs can be found in \citet{Kuerster2008}. 
In the RV analysis we used nightly averaged values of these measurements.
RVs presented in the table are with best fit RV offsets subtracted. For the four data sets these RV offsets are:
$\gamma$~UVES~=~$-$69.29$_{-1.44}^{+1.98}$ m\,s$^{-1}$, 
$\gamma$~FEROS~=~63159.42$_{-4.15}^{+4.79}$ m\,s$^{-1}$,
$\gamma$~HARPS~=~$-$334.29$_{-3.65}^{+2.60}$ m\,s$^{-1}$,
$\gamma$~HARPS+~=~$-$348.18$_{-3.21}^{+4.11}$ m\,s$^{-1}$.
 }

\end{table}

\end{document}